\documentclass[10pt,a4paper,twoside]{article}
\usepackage{epsfig}
\usepackage{baltlat6}
\usepackage{array}
\usepackage{here}
\begin{document}
\ \
\vspace{0.5mm}
\setcounter{page}{576}
\vspace{8mm}

\titlehead{Baltic Astronomy, vol.\,20, 576--579, 2011}

\titleb{STARK BROADENING OF SEVERAL Ar I SPECTRAL LINES IN THE VISIBLE PART OF THE SPECTRUM}

\begin{authorl}
\authorb{Milan S. Dimitrijevi\'c}{1,2},
\authorb{Andjelka Kova\v cevi\'c}{3},
\authorb{Zoran Simi\'c}{1} \\ and
\authorb{Sylvie Sahal-Br\'echot}{2}
\end{authorl}

\begin{addressl}
\addressb{1}{Astronomical Observatory, Volgina 7, 11060
Belgrade 38, Serbia;\\
mdimitrijevic@aob.bg.ac.rs, zsimic@aob.bg.ac.rs}
\addressb{2}{Observatoire de Paris, LERMA, 5 Place Jules Janssen, 92190 Meudon, France; sylvie.sahal-brechot@obspm.fr}
\addressb{3}{Department of Astronomy, Faculty of Mathematics, Studentski Trg 15, 11000  Belgrade, Serbia; andjelka@matf.bg.ac.rs}
\end{addressl}

\submitb{Received: 2011 August 8; accepted: 2011 August 15}

\begin{summary} In order to complete data
on Stark broadening parameters for Ar I line in the visible spectrum, we determined
Stark widths and shifts due to electron, proton, and ionized helium impacts,
for nine lines ($\lambda \lambda$ = 4191.0, 4259.4,
5912.1, 6043.2,  6045.0, 6752.9, 7503.9, 7514.6, 7724.2 $\AA$), using $jK$
coupling and semiclassical-perturbation theory.

The obtained results will enter the STARK-B database,
which is a part of Virtual Atomic and Molecular Data Center.
\end{summary}

\begin{keywords}
physical data and processes: Stark broadening, line profiles, databases
\end{keywords}

\resthead {Stark broadening of several Ar I spectral lines}
{M. S. Dimitrijevi\'c, A. Kova\v cevi\'c, Z. Simi\'c, S. Sahal-Br\'echot}

\sectionb{1}{INTRODUCTION}
Argon spectral lines, which
are useful not only for laboratory and technological plasmas but also
were observed in stellar atmospheres,
are of particular interest, especially in the visible spectrum. Stark broadening parameters of spectral lines are of interest for
diagnostics, modelling and investigations of various plasmas in
laboratory, technology, fusion research, laser designing, as well as for analysis, interpretation and synthesis
of Ar I lines in stellar spectra.

As the continuation of our previous work on the theoretical determination of Stark broadening of Ar I lines
(Dimitrijevi\'c et al., 2003, 2007, Milosavljevi\'c et al. 2003, Christova et al. 2010), and in order
to supplement the STARK-B database (http://stark-b.obspm.fr/), a part of Virtual Atomic and Molecular Data Center (VAMDC - http://vamdc.org/, Dubernet et al., 2010, Rixon et al., 2011), we determined here
Stark widths and shifts 
for nine lines at $\lambda \lambda$ = 4191.0, 4259.4,
5912.1, 6043.2,  6045.0, 6752.9, 7503.9, 7514.6, 7724.2 $\AA$ due to electron, proton, and ionized helium impacts, using the $jK$
coupling and semiclassical-perturbation theory (Sahal-Br\'echot 1969a,b, 1974, Dimitrijevi\'c \& Sahal-Br\'echot 1984).
 
\begin{table}[]
\begin{center}
\vbox{\footnotesize\tabcolsep=1pt
\parbox[c]{124mm}{\baselineskip=10pt
{\smallbf\ \ Table 1.}{\small\
 Stark full widths at half intensity maximum (W) and shifts (d) due to electron-, proton-, and ionized helium-impacts for Ar I, for a perturber density of 10$^{16}$ cm$^{-3}$. The quantity C (given in \AA cm$^{-3}$), when divided by the corresponding perturber width, gives an estimate for the maximum perturber density for which tabulated data may be used. The validity of the
impact approximation  has been estimated by checking if the collision volume (V) multiplied by the
perturber  density  (N)  is  much less   than   one (Sahal$-$Br\'
echot, 1969ab). With an asterisk are denoted values where 0.1 $<$
NV $\le $0.5.  \lstrut}}
\begin{tabular}{cccccccc}
\hline
      & &  Electrons&& Protons  & &           Ionized &helium \\
 TRANSITION &   T(K) &  W(\AA) & d(\AA) &  W(\AA) & d(\AA) &  W(\AA) & d(\AA)\hstrut\lstrut\\
\hline
 Ar I 4s -4p  & 2500.&  0.0761&     0.0512&  0.0333&  0.0129 & 0.0322&  0.0101\hstrut\\
$[3/2]_1-[1/2]_0$ & 5000.&  0.0851&     0.0620&  0.0341&  0.0149 & 0.0329&  0.0118\\
   7514.6 \AA   &10000.&  0.0967&     0.0610&  0.0350&  0.0171 & 0.0335&  0.0136\\
 C=0.27E+20   &20000.&  0.119 &     0.0586&  0.0361&  0.0195 & 0.0342&  0.0156\\
              &30000.&  0.135 &     0.0484&  0.0369&  0.0209 & 0.0347&  0.0168\\
              &50000.&  0.162 &     0.0389&  0.0380&  0.0229 & 0.0354&  0.0184\\
\hline
 Ar I 4s$'$-4p$'$ & 2500.&  0.0740&     0.0464&  0.0336&  0.0121 & 0.0328&  0.00951\\
$[1/2]_1-[1/2]_0$ & 5000.&  0.0820&     0.0578&  0.0343&  0.0140 & 0.0334&  0.0111\\
   7503.9 \AA   & 10000.& 0.0929&     0.0586&  0.0351&  0.0160 & 0.0339&  0.0128\\
 C=0.35E+20   &20000.&  0.115 &     0.0569&  0.0360&  0.0182 & 0.0345&  0.0146\\
              &30000.&  0.131 &     0.0481&  0.0367&  0.0196 & 0.0349&  0.0157\\
              &50000.&  0.157 &     0.0385&  0.0377&  0.0215 & 0.0355&  0.0172\\
\hline
 Ar I 4s$'$-4p$'$ & 2500.&  0.0665&     0.0379&  0.0307&  0.00961& 0.0301&  0.00759\\
$[1/2]_0-[1/2]_1$ & 5000.&  0.0713&     0.0458&  0.0312&  0.0111 & 0.0305&  0.00882\\
   7724.2 \AA   &10000.&  0.0795&     0.0442&  0.0316&  0.0127 & 0.0309&  0.0101\\
 C=0.44E+20   & 20000.& 0.0990&     0.0412&  0.0322&  0.0144 & 0.0312&  0.0115\\
              &30000.&  0.115 &     0.0342&  0.0327&  0.0155 & 0.0315&  0.0124\\
              &50000.&  0.141 &     0.0272&  0.0333&  0.0169 & 0.0319&  0.0136\\
\hline
 Ar I 4s$'$-5p$'$ & 2500.&  0.142 &     0.0941& *0.0524& *0.0218 &*0.0488& *0.0164\\
$[1/2]_1-[1/2]_0$ & 5000.&  0.164 &     0.113 & *0.0552& *0.0268 &*0.0517& *0.0207\\
   4259.4 \AA   &10000.&  0.186 &     0.128 &  0.0577&  0.0317 &*0.0538& *0.0249\\
 C=0.39E+19   &20000.&  0.212 &     0.125 &  0.0605&  0.0367 &*0.0558& *0.0292\\
              & 30000.& 0.229 &     0.109 &  0.0624&  0.0398 & 0.0571&  0.0317\\
              &50000.&  0.257 &     0.0908&  0.0651&  0.0439 & 0.0589&  0.0350\\
\hline
 Ar I 4s$'$-5p$'$ & 2500.&  0.117 &     0.0793& *0.0451& *0.0176 &*0.0424& *0.0134\\
$[1/2]_0-[3/2]_1$ & 5000.&  0.134 &     0.0893& *0.0470& *0.0214 &*0.0446& *0.0166\\
   4191.0 \AA   &10000.&  0.153 &     0.0997&  0.0488&  0.0252 &*0.0461& *0.0198\\
 C=0.40E+19   &20000.&  0.177 &     0.0962&  0.0509&  0.0291 &*0.0475& *0.0231\\
              &30000.&  0.194 &     0.0837&  0.0522&  0.0315 & 0.0484&  0.0251\\
              & 50000.& 0.220 &     0.0695&  0.0542&  0.0346 & 0.0497&  0.0277\\
\hline
 Ar I 4p -4d  & 2500.&  0.443 &     0.271 & *0.135 & *0.0591 &*0.124 &    *0.0443\\
$[1/2]_1-[3/2]_2$ & 5000.&  0.505 &     0.315 & *0.142 & *0.0730 &*0.132 &    *0.0565\\
   6752.9 \AA   &10000.&  0.585 &     0.315 &  0.150 &  0.0867 &*0.138 &    *0.0681\\
 C=0.60E+19   &20000.&  0.697 &     0.276 &  0.158 &  0.101  &*0.144 &    *0.0798\\
              &30000.&  0.773 &     0.241 &  0.163 &  0.109  & 0.148 &     0.0868\\
              &50000.&  0.862 &     0.202 &  0.171 &  0.120  & 0.153 &     0.0960\\
\hline

\end{tabular}
}
\end{center}
\end{table}

\begin{table}[]
\begin{center}
\vbox{\footnotesize\tabcolsep=1pt
\parbox[c]{124mm}{\baselineskip=10pt
{\smallbf\ \ Table 1.}{\small\
 Continued    \lstrut}}
\begin{tabular}{cccccccc}
\hline
      & &  Electrons&& Protons  & &           Ionized &helium \\
 TRANSITION &   T(K) &  W(\AA) & d(\AA) &  W(\AA) & d(\AA) &  W(\AA) & d(\AA)\hstrut\lstrut\\
\hline
 
 Ar I 4p -4d$'$ & 2500.&  0.650 &     0.430 & *0.162 & *0.0856 &       &\\
$[1/2]_1-[3/2]_1$ & 5000.&  0.721 &     0.495 & *0.178 & *0.112  &       &  \\
   5912.1 \AA   &10000.&  0.778 &     0.485 & *0.194 & *0.138  &*0.167 &    *0.107\\
 C=0.22E+19   &20000.&  0.843 &     0.432 & *0.212 & *0.163  &*0.181 &    *0.128\\
              &30000.&  0.887 &     0.382 & *0.224 & *0.178  &*0.190 &    *0.141\\
              &50000.&  0.937 &     0.309 &  0.240 &  0.197  &*0.202 &    *0.157\\
\hline
 Ar I 4p -5d  & 2500.&   1.76 &      1.07 &        &         &       &            \\
$[5/2]_2-[7/2]_3$ & 5000.&   1.91 &     0.919 &        &         &       &            \\
   6043.2 \AA   &10000.&   2.24 &     0.667 &        &         &       &            \\
 C=0.11E+19   &20000.&   2.53 &     0.504 &        &         &       &            \\
              &30000.&   2.60 &     0.376 & *0.636 & *0.487  &       &            \\
              &50000.&   2.67 &     0.213 & *0.690 & *0.547  &       &            \\
\hline
 Ar I 4p$'$-6d  & 2500.& * 4.89 &    * 2.86          &         &       &            \\
$[3/2]_2-[5/2]_3$ & 5000.& * 5.31 &    * 3.14          &         &       &            \\
   6045.0 \AA   &10000.& * 5.73 &    * 3.01          &         &       &            \\
 C=0.80E+18   &20000.&   6.32 &      2.36          &         &       &            \\
              &30000.&   6.61 &      1.92          &         &       &            \\
              &50000.&   6.78 &      1.48          &         &       &            \lstrut\\
\hline

\end{tabular}
}
\end{center}
\end{table}

\sectionb{2}{RESULTS AND DISCUSSION}

Stark broadening parameters for nine Ar I lines have been determined within the semiclassical perturbation formalism, discussed in detail in Sahal-Br\'echot (1969ab). The optimization and updates can be found in e.g. Sahal-Br\'echot (1974), Dimitrijevi\'c \& Sahal-Br\'echot (1984).
Energy levels have been taken from Bashkin \& Stoner (1978).
The obtained results for electron-, proton, and ionized helium-impact widths (FWHM) and shifts are shown  in
Table 1,
for a perturber density of  10$^{16}$ cm$^{-3}$ and
temperatures between 2500 and 50000 K. For perturber densities lower
than tabulated, Stark broadening parameters may be scaled linearly.

For each   value   given   in   Table   1,  the collision
volume (V) multiplied by the perturber  density  (N)  is  much
less   than   one  and   the  impact  approximation  is  valid
(Sahal-Br\'echot 1969a).
When   the   impact  approximation  is  not  valid,  the   ion
broadening  contribution   may   be    estimated    by   using
quasistatic approach (Griem 1974, or Sahal-Br\'echot 1991).
In the region between where neither of these two approximations is
valid, a unified type theory should be used.

The
accuracy of the results obtained decreases when broadening by ion
interactions becomes important.

For extrapolation of values from Table 1 towards higher electron densities, the results start to deviate from linear behavior due to
Debye screening. This effect is more important for the shift than for
the width and could be taken into account by the approximative method described in Griem (1974).

The obtained results will enter in the STARK-B database (http://stark-b.obs
pm.fr/), which is a part of Virtual Atomic and Molecular Data Center (VAMDC - http://vamdc.org/, Dubernet et al. 2010, Rixon et al., 2011).
supported by EU
in the framework of the FP7 "Research Infrastructures - INFRA-2008-1.2.2 -
Scientific Data Infrastructures" initiative, with aim to build an interoperable
e-Infrastructure for the exchange of atomic and molecular data.

\thanks{ This work has been supported by VAMDC,  funded under the "Combination of Collaborative Projects and Coordination and  Support Actions" Funding Scheme of The Seventh Framework Program. Call topic: INFRA-2008-1.2.2 Scientific Data Infrastructure. Grant Agreement number: \break
239108.  The authors are also  grateful for the support  provided by Ministry Education and Science of Republic of Serbia through project  176002 "Influence of collisional processes on astrophysical plasma spectra".}

\References

\refb  Bashkin S.,  Stoner Jr. J. J. 1978, Atomic Energy Levels
and Grotrian Diagrams, Vol. 2, North Holland, Amsterdam

\refb Christova, M., Dimitrijevi\'c M. S., Kova\v cevi\'c, A. 2010, J. Phys. Conf. Ser., 207, 012024

\refb Dimitrijevi\'c M.S., Christova M., Sahal-Br\'echot S. 2007, 	Phys. Scr., 75, 809

\refb Dimitrijevi\'c M. S., Skuljan, Lj., Djeni\v ze S. 2003,  Phys. Scr., 66, 77

\refb Dimitrijevi\'c M. S., Sahal-Br\'echot S. 1984, JQSRT, 31, 301

\refb Dubernet M. L. et al. 2010, JQSRT, 111, 2151

\refb Griem H. R., 1974, Spectral Line Broadening by
Plasmas, Academic Press, New York

\refb Milosavljevi\'c V., Dimitrijevi\'c M. S., Djeni\v ze S. 2003, High Temp. Material Processes, 7, 525

\refb  Rixon G., Dubernet M. L., Piskunov N. et al., 2011, AIP Conference Proceedings 1344, 107

\refb Sahal-Br\'echot S. 1969a, A\&A, 1, 91

\refb Sahal-Br\'echot S. 1969b, A\&A, 2, 322

\refb Sahal-Br\'echot S. 1974, A\&A, 35, 321

\refb Sahal-Br\'echot S. 1991, A\&A, 245, 322
\end{document}